\documentclass[sigconf,authorversion,nonacm]{acmart}
\AtBeginDocument{%
  }

\usepackage{url}
\usepackage{subcaption}
\usepackage{enumitem}
\usepackage{graphicx}
\usepackage{tabularx}
\usepackage{multirow}
\usepackage{listings}
\usepackage{adjustbox}
\usepackage{calc}
\usepackage{xspace}
\usepackage{pdflscape}
\usepackage{tcolorbox}

\newcounter{finding_counter}
\newcommand{\boxtext}[1]{
    \begin{tcolorbox}[flushleft upper,boxrule=1pt,arc=1pt,left=0pt,right=0pt,top=0pt,bottom=0pt,colback=white,after=\ignorespacesafterend\par\noindent]
        \textbf{Observation \arabic{finding_counter}:} #1 \stepcounter{finding_counter}
    \end{tcolorbox}
}
\newcommand{\bl}{{BL}\xspace}
\newcommand{\electra}{{ELECTRA}\xspace}

\date{February 2024}

\begin{document}
\title[Aligning Programming Language and Natural Language]{Aligning Programming Language and Natural Language: Exploring Design Choices in Multi-Modal Transformer-Based Embedding for Bug Localization}
\begin{abstract}

Bug localization refers to the identification of source code files which is in a programming language and also responsible for the unexpected behavior of software using the bug report, which is a natural language. As bug localization is labor-intensive, bug localization models are employed to assist software developers. Due to the domain difference between source code files and bug reports, modern bug-localization systems, based on deep learning models, rely heavily on embedding techniques that project bug reports and source code files into a shared vector space. The creation of an embedding involves several design choices, but the impact of these choices on the quality of embedding and the performance of bug localization models remains unexplained in current research.

To address this gap, our study evaluated 14 distinct embedding models to gain insights into the effects of various design choices. Subsequently, we developed bug localization models utilizing these embedding models to assess the influence of these choices on the performance of the localization models. Our findings indicate that the pre-training strategies significantly affect the quality of the embedding. Moreover, we discovered that the familiarity of the embedding models with the data has a notable impact on the bug localization model's performance. Notably, when the training and testing data are collected from different projects, the performance of the bug localization models exhibits substantial fluctuations.

\end{abstract}
\author{Partha Chakraborty}
\affiliation{%
  \institution{University of Waterloo}
  \city{Waterloo}
  \state{Ontario}
  \country{Canada}}
\email{p9chakra@uwaterloo.ca}

\author{Venkatraman Arumugam}
\affiliation{%
  \institution{University of Waterloo}
  \city{Waterloo}
  \state{Ontario}
  \country{Canada}}
\email{venkatraman.arumugam@uwaterloo.ca}

\author{Meiyappan Nagappan}
\affiliation{%
  \institution{University of Waterloo}
  \city{Waterloo}
  \state{Ontario}
  \country{Canada}}
\email{mei.nagappan@uwaterloo.ca}

\maketitle
\section{Introduction}
\label{sec-introduction}

In the field of software engineering, a bug is considered a deviation from the ideal behavior. The first step in fixing a bug is the identification of its location in the codebase. There have been many studies that attempt to automate the process. The studies proposed techniques based on Information Retrieval (IR)~\citep{Saha2013,Wang2014,Wen2016,Wong2014,Youm2015,Zhou2012} or using Machine Learning (ML)~\citep{Akbar2020,Moin2010,Liu2005,Le2015,Dallmeier2005} or Deep Learning (DL)~\citep{Huo2019,huo2016learning,Wang2020,Jiang2020,Zhu2020,Xiao2018}.

DL-based techniques have achieved comparatively higher performance than other techniques in recent years. Most DL-based techniques use similarity measurement to identify buggy files. However, the bug reports are in Natural Language (NL), and source codes are in Programming Language (PL). Thus, there is a semantic and lexical gap between them. Previous studies used very well-known NLP techniques such as  Word2Vec or FastText to project the source code and the bug report in the same vector space. This vector representation was successful to some extent in bridging the beforementioned gap. However, one of the drawbacks of such techniques is that they do not consider the context in generating vector representation of a source code file or a bug report. In recent times state-of-the-art large pre-trained models such as BERT~\citep{devlin2018bert}, RoBERTa~\citep{liu2019} are used in Natural Language Processing (NLP) for capturing the semantic and contextual representation of text as they consider the context in generating vector representation. The advantage of using those models is that they are more context-sensitive than before.

\begin{figure}[!tb] 
 \centering
 \includegraphics[scale=0.45]{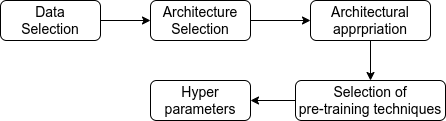}
 \caption{Steps of training a transformer.}
 \label{fig:Steps}
\end{figure}

However, there are too many design choices in creating a transformer-based embedding, such as model architecture, pre-training strategies, and data sources for training the embedding model. Figure~\ref{fig:Steps} presents the steps of training a transformer. These design choices can impact the quality of the embeddings generated by an embedding model. In the first step of training transformers, we have to identify the appropriate data for pre-training. In the second and third steps, we have to select the proper architecture of the transformers and make the appropriate changes to fit the use case (e.g., sequence size, attention head). In the third step, we have to identify the pre-training technique. So far, there are different pre-training techniques such as Masked Language Model (MLM), Question-Answering (QA), Next Sentence Prediction (NSP), etc. However, the performance of the pre-training technique is specific to the use case~\citep{liu2019}. The final step is the selection of hyper-parameters such as learning rate and batch size. Onta{\~{n}}{\'{o}}n et al.~\citep{Santiago2021} have focussed on the second step and explored the impact of different types of positional encodings, different types of decoders and weight sharing. Zhu et al.~\citep{Zhu2021} have also focussed on identifying the impact of design choices in the second step. In their study, they used reinforcement learning to identify the best design choices, for example, the number of layers in the transformer model and the type of activation function. The performance of the bug localization model is dependent on the embedding quality. There are too many types of architectures and training methods for transformer-based models. Thus, knowing which architecture and training methods produce the best quality embedding for bug localization task is important.\par

With the increment of the complexity of the modern bug -localization model, the resource (GPU size, training time) requirement is also increasing. Thus, a project-specific bug localization model is a less popular choice than a cross-project bug localization model, which can localize bugs in different projects. However, the performance of cross-project bug localization models is comparatively lower than project-specific bug localization models. Thus, it is required to quantify the impact of project-specific data on bug localization models' performance to understand the trade-off and make better design decisions.\par

This study aims to understand the impact of three design choices on embedding models' performance and the generalization capability of those embedding models. The choices are, the use of domain-specific data, pre-training methodology, and the sequence length of the embedding. Though the design space for a language model (embedding) is not limited, we can say that the type of training data, model architecture, and training strategies are the primary design choices. Prior studies have already identified some other design choices such as learning rate, weight sharing~\citep{Santiago2021}, and other hyperparameters~\citep{Izsak2021,Quijano2021}. Thus in this study, we intended to identify the impact of three design choices by answering the following research questions.

\begin{enumerate}[label={\textbf{RQ\arabic{*}.}}, leftmargin=15pt]
    \item \textbf{Do we need data familiarity to apply the embeddings?}\\
    In this research question, we will test whether project-specific data is needed for the embedding models or not. In the NLP domain, we have observed the use of transfer learning. However, previous studies in the domain of bug localization used project-specific embedding~\citep{Liang2022} for their models. Using project-specific embedding may not be a viable approach in commercial settings. Thus we need to verify whether data familiarity is required in using transformer-based embedding models. We trained embedding models on two different datasets to answer the question. We found that pre-trained embedding models using project-specific datasets perform better in bug localization tasks than those that are not pre-trained using those data.
    
    \item \textbf{Do pre-training methodologies impact embedding models performance?}\\
     We will analyze whether pre-training impacts the performance of the embedding model and which pre-training technique is better for bug localization in this research question. We have seen different pre-training techniques such as MLM, QA, NSP, etc. However, pre-training techniques have domain-specific use. Peter et al.~\citep{peters-etal-2019-tune} found that feature extraction depends on the similarity of the pre-training and target tasks. Liu et al.~\citep{liu2019} found that the individual sentence NSP (Next Sentence Prediction) task hurts the performance of the transformer model in a question-answer task. No other studies before us have verified the effectiveness of the pre-training techniques in the domain of bug localization which uses both programming language and natural language. Thus we need to know the impact of pre-training techniques and identify the best technique for bug localization tasks. For this, we pre-trained the embedding models using several pre-training strategies. We found that certain pre-training methodologies, such as \electra (Efficiently Learning an Encoder that Classifies Token Replacements Accurately) create better embedding models than others.
    
    \item \textbf{Do transformers with longer maximum input sequence lengths perform well compared to the models with shorter maximum input sequences?}\\
    We test whether the length of the input sequences of an embedding model impacts the model's performance in the bug localization task. Source codes are typically long ($\sim$2000 tokens), whereas transformers typically support 512 tokens. Typical transformer architectures can not be extended to higher sequence lengths as the action is computationally expensive($O(n^4)$, where n is sequence length~\cite{beltagy2020longformer}). Thus we need to understand the performance gain over using a long input sequence transformer in bug localization tasks. We trained embedding models with different lengths and compared their performance in the bug-localization task. In most experimental settings we conducted, the embedding model with longer input sequences outperformed the embedding model with shorter input sequences.  
\end{enumerate}

We used two datasets to train the embedding and bug localization models. We collected the first one by mining the bugs/issues from the Apache project and Github. The other one has been offered by Ye et al.~\citep{Ye2014}. We have evaluated 14 transformer-based embedding models on the bug-localization task using those two datasets.

Overall, our study finds a significant difference between the in-project and cross-project performance of the embedding models. We also found that pre-training methodologies impact embedding models' performance. Our research found that in most cases, \electra pre-trained models performed better than others. This study has trained embedding models using bug report-source code pair. Compared to the documentation-source code pair-trained model, we found bug report-source code-trained embedding models performed better. The dataset and code used in this study are publicly available here~\footnote{\url{https://zenodo.org/record/6760333}\label{rep_package}}.
\section{Dataset}
\label{sec-dataset}

We have used two different datasets for this study. The first one consists of mined projects of the Apache project and public Github repositories. The following steps are followed to extract data from JIRA (bug report repository of Apache projects).
\begin{enumerate}[left=0.3pt]
    \item \textbf{Candidate project selection:} In this step, we listed all Apache projects except those included in the dataset prepared by  Ye et al.~\citep{Ye2014} or Bugzbook~\citep{Akbar2020}. The reason for the exclusion is that we intend to use those datasets to test the proposed embedding's effectiveness in the future.
    
    \item \textbf{Extraction of bug reports:} After creating the candidate project list, we have extracted all the fixed bug reports. For that purpose, we have used Jira Query Language (JQL). However, not all Apache projects selected in the previous step have enough bug reports. We have filtered out the project if that has less than ten bug reports in JIRA. The list of the projects is available in the online Appendix~\footref{rep_package}.
    
    \item\textbf{Extraction of source code:} In this step, each bug report in JIRA, identifiable by a unique id, was linked to its corresponding pull request in Git/Github using the bug id and a regular expression heuristic, as done in prior studies~\citep{liwerski2005}. Following this, the commit hash id (SHA) of the fix, along with the pre-and post-fix file versions for each change in the commit, were extracted.
\end{enumerate}

For public repositories of Github, first, we listed all repositories of Java language sorted by the number of stars in descending order. Like the steps followed for extracting JIRA data, we also excluded some of the repositories in this step. After that, we followed the before-mentioned steps to extract bug reports and source codes from those repositories.

For this study, we have used only the Java language source code files and bug reports. After filtering by language, we have 7,970 bug reports from 21 projects. From now on, we will refer to this dataset as  Bug Localization Dataset (BLDS). This dataset has been used only for pre-training the language model.

The second dataset we used has been offered by Ye et al.~\citep{Ye2014, Ye2014_dataset}. This dataset contains 22.7K bug reports from Six popular Apache projects (AspectJ, Birt, Eclipse UI, JDT, SWT, and  Tomcat) and associated source codes. From now on, we will refer to this dataset as the Benchmark Dataset for Bug Localization (Bench-BLDS). We select this dataset as prior study~\cite{Kim2021} showed that this dataset contains the lowest number of false positive or negative cases among other datasets~\cite{Lee2018, Moreno2015} in the literature. We have used Bench-BLDS only for training and testing the bug localization model. The Bench-BLDS dataset contains 12480 bug reports from six projects. Due to space restriction, the length distribution of the source code files of BLDS and Bench-BLDS along with the description of the projects is available in the online Appendix~\footref{rep_package}.

\section{Methodology}
\label{sec-methodology}
\begin{figure}[!tb] 
 \centering
 \includegraphics[scale=0.35]{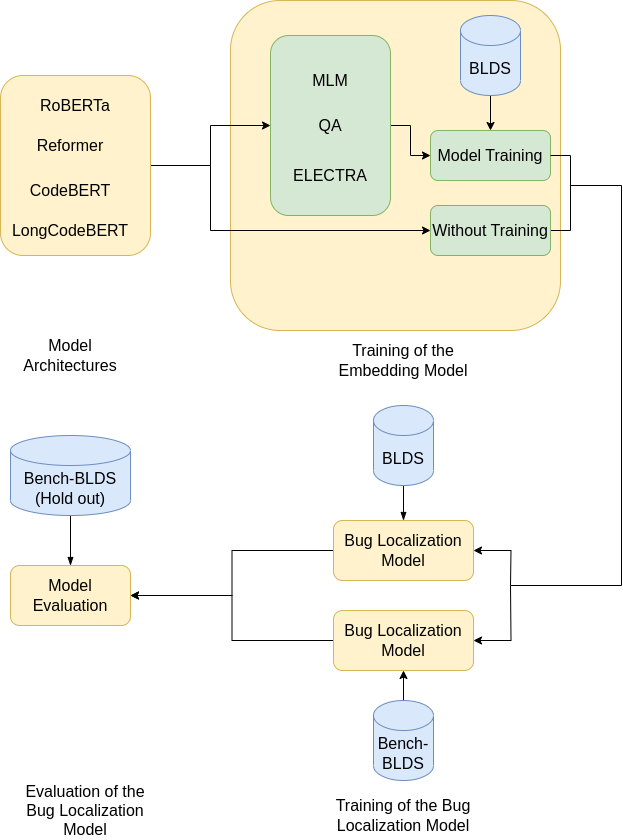}
 \caption{Multi-modal embedding training pipeline.}
 \label{fig:methodology}
\end{figure}
In this study, we created 12 different transformer-based embedding models. Previous studies have presented one multi-modal language model~\citep{Feng2020}, and we have updated that language model by extending the maximum supported sequence length. Next, we evaluated 14 models (our twelve models along with one model from the previous study and the extended version of that model) in the bug localization task. The embedding models differ from each other in terms of the architecture and the pre-training methodology. The methodology of this study has been presented in Figure~\ref{fig:methodology}.\par

\noindent\textbf{Model Architectures. } The first step of our methodology is the selection of the architecture for the embedding model. We have used two different architectures RoBERTa, and Reformer. The reason for selecting these two embedding models is the methodology for calculating attention is fundamentally differrent for these two architectures. RoBERTa is a popular architecture; based on this architecture, many domain-specific embedding models have been proposed in NLP. On the other hand, Reformer presented a new attention calculation method that reduces the complexity of attention calculation. The weights of the models are initialized randomly. Besides these two architectures, we have also used CodeBERT~\cite{Feng2020}, which is a variant of RoBERTa model trained on multimodal (NL-PL) data. As CodeBERT has already been trained, we have used the trained weights for CodeBERT. Furthermore, we extended the CodeBERT model by increasing the maximum sequence length. In this process, we have followed the approach of Beltagy et al.~\citep{beltagy2020longformer} The extended model will have a new maximum sequence length while using the trained weights of CodeBERT.  From now on, we will call this extended CodeBERT model LongCodeBERT. We have selected CodeBERT~\citep{Feng2020} as it is the state-of-the-art multi-modal embedding model for software engineering tasks. The reason behind using LongCodeBERT is it extends the maximum sequence length of the original CodeBERT embedding model while not changing the weights. After this step, we have four different architectures of the embedding model, two of which (CodeBERT, LongCodeBERT) have already been trained in a previous study.\par
\noindent{\textbf{Training of the Embedding Model. } The second step is the pre-training of the embedding models. For pre-training the embedding models, we followed three methods: Masked Language Modeling (MLM)~\cite{devlin2018bert}, \electra ~\cite{clark2020electra}, and QA~\cite{Glass2020}, and used the BLDS dataset. After this step, we will have twelve different embedding models (four architecture trained using three different methods). Moreover, we also wanted to know the performance of the embedding models proposed in previous studies. Thus, we have included the CodeBERT and LongCodeBERT embedding models without further fine-tuning/pre-training steps, which increases the number of embedding models from twelve to fourteen (four architecture trained using three different methods along with two architectures without any further training).\par
\noindent\textbf{Training of the Bug Localization Model. }
In this study, we referred to the language models (DL models that generate multi-modal data embedding) as \emph{embedding models} and bug detection models (DL models that use embedding models to detect bugs) as \ emph{bug localization models}. In the third step, we have used the embeddings generated from the embedding models to train bug localization model. Following the practices of previous studies~\citep{Huo2021, Wang2020, huo2016learning, Xiao2018}, we have used a Convolutional Neural Network (CNN) based architecture for the bug localization model. This study aims not to find the best bug localization model but the best embedding training technique for a bug localization model. We have used the bug localization models as a performance measurement for the embedding models. The bug localization model consists of three convolutional layers followed by a perceptron layer. In training the bug-localization models, we have used two different datasets. The input to the bug localization models is a vector of the bug report and the source code corresponding to the bug report, concatenated together. The bug localization task has been aimed for file-level bug localization and has narrowed down to the task of a binary classification problem. The model is trained to learn whether the given pair of bug reports and code snippet is a match or not. All the layers/weights of the embedding model are unchanged (frozen) while training the bug-localization model. After this step, we will have two bug localization models for each of the fourteen embedding models, which means we will have 28 bug localization models.\par
\noindent\textbf{Evaluation of the Bug Localization Model.} In the fourth step, we evaluated the bug localization models' performance on a held-out test dataset created from the Bench-BLDS. The performance has been measured in terms of Mean Reciprocal Rank (MRR). We have mentioned before that BLDS and Bench-BLDS dataset has no common projects. Thus, the model trained on the BLDS dataset and evaluated on the Bench-BLDS dataset can be considered a cross-project bug localization model. All the models are trained for ten epochs in our training setup with 32 samples per GPU separately. Later the performance is evaluated on the held out Bench-BLDS dataset.

\noindent We have used all combinations of model architectures, training techniques, and data sources to understand the impact and identify the best design choices. However, testing all combinations requires time. For example, to complete one row of Table~\ref{table: average_model} requires ~sixteen days of training in a single GPU machine (Nvidia V100 Volta 32G GPU). This translates to almost 2 months for the entire set of combinations reported in Table 1. Now any small tweak that we make, we need to rerun the whole experiment again which would take anywhere between 4-56 days. Thus another contribution of this study is that future studies do not have to go through the same time-consuming process and test all combinations to find good design choices for embeddings.
\section{Results}
\label{sec-results}
\begin{table}[!tb]
\centering
\caption{Performance of the IR-based systems.}
\begin{tabular}{|cccc|}
\hline
\multicolumn{1}{|c|}{\textbf{Study}} & \multicolumn{1}{c|}{\textbf{AspectJ}} & \multicolumn{1}{c|}{\textbf{JDT}} & \textbf{SWT} \\ \hline
\multicolumn{4}{|c|}{\textbf{MRR}}                                                                                              \\ \hline
\multicolumn{1}{|c|}{BugLocator~\citep{Zhou2012}}     & \multicolumn{1}{c|}{0.38}             & \multicolumn{1}{c|}{0.26}         & 0.50         \\ \hline
\multicolumn{1}{|c|}{BRTracer~\citep{Wong2014}}       & \multicolumn{1}{c|}{0.25}             & \multicolumn{1}{c|}{0.33}         & 0.62         \\ \hline
\multicolumn{1}{|c|}{BLUiR~\citep{Saha2013}}          & \multicolumn{1}{c|}{0.41}           & \multicolumn{1}{c|}{0.38}         & 0.59         \\ \hline
\multicolumn{1}{|c|}{AmaLgam~\citep{Wang2014}}        & \multicolumn{1}{c|}{0.33}             & \multicolumn{1}{c|}{0.33}         & 0.62         \\ \hline
\multicolumn{4}{|c|}{\textbf{MAP}}                                                                                              \\ \hline
\multicolumn{1}{|c|}{BugLocator}     & \multicolumn{1}{c|}{0.21}             & \multicolumn{1}{c|}{0.16}         & 0.44         \\ \hline
\multicolumn{1}{|c|}{BRTracer}       & \multicolumn{1}{c|}{0.14}             & \multicolumn{1}{c|}{0.24}         & 0.55         \\ \hline
\multicolumn{1}{|c|}{BLUiR}          & \multicolumn{1}{c|}{0.22}             & \multicolumn{1}{c|}{0.27}         & 0.52         \\ \hline
\multicolumn{1}{|c|}{AmaLgam}        & \multicolumn{1}{c|}{0.20}             & \multicolumn{1}{c|}{0.24}         & 0.55         \\ \hline
\end{tabular}
\label{table:ir_performance}
\end{table}
\begin{table*}[!htb]
\centering
\caption{Average performance with varying data familiarity.}
\begin{tabular}{|rrrrrrrr|}
\hline
\multicolumn{1}{|c|}{\textbf{\begin{tabular}[c]{@{}c@{}}Training\\ Data (CNN Model)\end{tabular}}} & \multicolumn{1}{c|}{\textbf{AspectJ}} & \multicolumn{1}{c|}{\textbf{Birt}} & \multicolumn{1}{c|}{\textbf{Eclipse}} & \multicolumn{1}{c|}{\textbf{JDT}} & \multicolumn{1}{c|}{\textbf{SWT}} & \multicolumn{1}{c|}{\textbf{Tomcat}} & \multicolumn{1}{c|}{\textbf{Overall}} \\ \hline
\multicolumn{8}{|c|}{\textbf{MRR}}                                                                                                                                                                                                                                                                                                                                             \\ \hline
\multicolumn{1}{|r|}{BLDS}                                                                         & \multicolumn{1}{r|}{0.28}             & \multicolumn{1}{r|}{0.26}          & \multicolumn{1}{r|}{0.28}             & \multicolumn{1}{r|}{0.27}         & \multicolumn{1}{r|}{0.25}         & \multicolumn{1}{r|}{0.25}            & 0.27                                  \\ \hline
\multicolumn{1}{|r|}{Bench-BLDS}                                                                   & \multicolumn{1}{r|}{0.37}             & \multicolumn{1}{r|}{0.32}          & \multicolumn{1}{r|}{0.37}             & \multicolumn{1}{r|}{0.37}         & \multicolumn{1}{r|}{0.35}         & \multicolumn{1}{r|}{0.30}            & 0.35                                  \\ \hline
\multicolumn{8}{|c|}{\textbf{MAP}}                                                                                                                                                                                                                                                                                                                                             \\ \hline
\multicolumn{1}{|r|}{BLDS}                                                                         & \multicolumn{1}{r|}{0.19}             & \multicolumn{1}{r|}{0.18}          & \multicolumn{1}{r|}{0.20}             & \multicolumn{1}{r|}{0.18}         & \multicolumn{1}{r|}{0.17}         & \multicolumn{1}{r|}{0.16}            & 0.17                                  \\ \hline
\multicolumn{1}{|r|}{Bench-BLDS}                                                                   & \multicolumn{1}{r|}{0.28}             & \multicolumn{1}{r|}{0.23}          & \multicolumn{1}{r|}{0.29}             & \multicolumn{1}{r|}{0.27}         & \multicolumn{1}{r|}{0.27}         & \multicolumn{1}{r|}{0.21}            & 0.25                                  \\ \hline
\end{tabular}
\label{table: average_data_familiarity}
\end{table*}
\begin{table*}[!tb]
\centering
\caption{Average performance in different pre-training techniques.}
\begin{tabular}{|rrrrrrrr|}
\hline
\multicolumn{1}{|c|}{\textbf{\begin{tabular}[c]{@{}c@{}}Embedding Training\\ Strategy\end{tabular}}} & \multicolumn{1}{c|}{\textbf{AspectJ}} & \multicolumn{1}{c|}{\textbf{Birt}} & \multicolumn{1}{c|}{\textbf{Eclipse}} & \multicolumn{1}{c|}{\textbf{JDT}} & \multicolumn{1}{c|}{\textbf{SWT}} & \multicolumn{1}{c|}{\textbf{Tomcat}} & \multicolumn{1}{c|}{\textbf{Overall}} \\ \hline
\multicolumn{8}{|c|}{\textbf{MRR}}                                                                                                                                                                                                                                                                                                                                               \\ \hline
\multicolumn{1}{|r|}{ \electra}                                                                        & \multicolumn{1}{r|}{0.37}             & \multicolumn{1}{r|}{0.28}          & \multicolumn{1}{r|}{0.34}             & \multicolumn{1}{r|}{0.35}         & \multicolumn{1}{r|}{0.33}         & \multicolumn{1}{r|}{0.30}            & 0.33                                  \\ \hline
\multicolumn{1}{|r|}{MLM}                                                                            & \multicolumn{1}{r|}{0.36}             & \multicolumn{1}{r|}{0.29}          & \multicolumn{1}{r|}{0.34}             & \multicolumn{1}{r|}{0.37}         & \multicolumn{1}{r|}{0.31}         & \multicolumn{1}{r|}{0.28}            & 0.33                                  \\ \hline
\multicolumn{1}{|r|}{MLM and QA}                                                                     & \multicolumn{1}{r|}{0.27}             & \multicolumn{1}{r|}{0.30}          & \multicolumn{1}{r|}{0.32}             & \multicolumn{1}{r|}{0.28}         & \multicolumn{1}{r|}{0.27}         & \multicolumn{1}{r|}{0.27}            & 0.29                                  \\ \hline
\multicolumn{1}{|r|}{Without training}                                                               & \multicolumn{1}{r|}{0.27}             & \multicolumn{1}{r|}{0.27}          & \multicolumn{1}{r|}{0.28}             & \multicolumn{1}{r|}{0.23}         & \multicolumn{1}{r|}{0.27}         & \multicolumn{1}{r|}{0.25}            & 0.27                                  \\ \hline
\multicolumn{8}{|c|}{\textbf{MAP}}                                                                                                                                                                                                                                                                                                                                               \\ \hline
\multicolumn{1}{|r|}{ \electra}                                                                        & \multicolumn{1}{r|}{0.29}             & \multicolumn{1}{r|}{0.20}          & \multicolumn{1}{r|}{0.26}             & \multicolumn{1}{r|}{0.26}         & \multicolumn{1}{r|}{0.25}         & \multicolumn{1}{r|}{0.21}            & 0.23                                  \\ \hline
\multicolumn{1}{|r|}{MLM}                                                                            & \multicolumn{1}{r|}{0.27}             & \multicolumn{1}{r|}{0.21}          & \multicolumn{1}{r|}{0.25}             & \multicolumn{1}{r|}{0.28}         & \multicolumn{1}{r|}{0.23}         & \multicolumn{1}{r|}{0.19}            & 0.23                                  \\ \hline
\multicolumn{1}{|r|}{MLM and QA}                                                                     & \multicolumn{1}{r|}{0.19}             & \multicolumn{1}{r|}{0.22}          & \multicolumn{1}{r|}{0.24}             & \multicolumn{1}{r|}{0.19}         & \multicolumn{1}{r|}{0.19}         & \multicolumn{1}{r|}{0.18}            & 0.19                                  \\ \hline
\multicolumn{1}{|r|}{Without training}                                                               & \multicolumn{1}{r|}{0.18}             & \multicolumn{1}{r|}{0.18}          & \multicolumn{1}{r|}{0.20}             & \multicolumn{1}{r|}{0.14}         & \multicolumn{1}{r|}{0.19}         & \multicolumn{1}{r|}{0.16}            & 0.17                                  \\ \hline
\end{tabular}
\label{table: average_pretrain}
\end{table*}
Table ~\ref{table: average_data_familiarity}, ~\ref{table: average_pretrain} and~\ref{table: average_model} represents the average performance whether the \bl model is trained on project-specific data or not, the average performance of each pre-training technique, and  average performance of each architecture respectively. Each column with the project name (AspectJ, Birt, JDT, SWT, and Tomcat) represents the models' performance in those respective projects. The \emph{overall} column represents the overall performance of the model in these six projects. Table ~\ref{table: average_data_familiarity}, ~\ref{table: average_pretrain} and~\ref{table: average_model} have been calculated from Table A.2 and Table A.3, which are available in the online appendix~\cite{rep_package}. Tables A.2 and A.3 represent the performance of all the bug localization models in terms of MRR and MAP, respectively. Besides, we have also presented the performance of previous IR-based studies in Table~\ref{table:ir_performance}. However, those studies have not used Birt, Eclipse UI, and Tomcat. Thus, the performance of the tools on those projects was not presented.
\subsection{RQ1. Do we need data familiarity to apply the embeddings?}
\label{RQ1}
In the natural language domain, pre-trained embedding models are used in various downstream tasks such as question-answering, sentiment detection, etc. To use pre-trained embedding models in downstream tasks, one needs to add a task-specific model (head) on top of the embedding models. From a language model, we only receive the vector representation of the text in the embedding space. However, in the context of source code, the use of libraries, comment style, the coding style is not the same in all projects. 
Thus, the relation between vector representation and bugginess may vary from project to project. This variation may pose a problem if we intend to use a bug localization model in another project without further training. The head is not familiar with the relation between the vector of the source code-bug report pair and the bugginess of a file. Thus, the performance may vary in a new project.\par
Let's think about bug localization in commercial settings where a tool is used to localize bugs from hundreds of different projects of an organization. Project-specific training seems like a less viable approach. It will require a high amount of resources to maintain (train and deploy) a specific head for each project. The scalability of a bug localization system depends on whether project-specific training is required or not. Past NLP research~\citep{howard2018} seems to indicate that transformer-based models are good at predicting when they are not pre-trained on similar data. Thus, in this research question, we will test whether or not project-specific training is necessary for the heads.\par 
Table~\ref{table: average_data_familiarity} represents the average performance in two cases, whether or not the \bl model is trained on project-specific data. From Table~\ref{table: average_data_familiarity} we can observe that the models trained on the Bench-BLDS dataset performed better than the model trained on the BLDS dataset. To verify whether the observation is statistically significant, we conducted a pairwise Mann-Whitney test between the model trained on BLDS and Bench-BLDS. We found that the observation that Bench-BLDS trained models perform better is statistically significant ($p<0.001$) for both MRR and MAP. The observation may point out that the embedding model used in the understanding of programming language is learning more project-specific features than the ones in the NLP domain. For example, in the text classification task using of ULMFit~\citep{howard2018} without pre-training achieved a 5.63\% error, or the pre-trained embedding model with project-specific data achieved a 5.00\% error on the IMDB dataset. However, in this study, we have observed a maximum 76\% drop in performance. From a high-level understanding, we can say that the heads trained on specific data representations struggle with generalization across projects, indicating a need for more adaptable architecture for heads.
\boxtext{Data familiarity has an impact on embedding models' performance. Training the embeddings with project-specific data (fine-tuning) can enhance the performance of the bug localization models.}
\begin{table*}[!tb]
\centering
\caption{Average performance with varying model architecture.}
\begin{tabular}{|rrrrrrrr|}
\hline
\multicolumn{1}{|l|}{\textbf{\begin{tabular}[c]{@{}l@{}}Model\\ Name\end{tabular}}} & \multicolumn{1}{l|}{\textbf{AspectJ}} & \multicolumn{1}{l|}{\textbf{Birt}} & \multicolumn{1}{l|}{\textbf{Eclipse}} & \multicolumn{1}{l|}{\textbf{JDT}} & \multicolumn{1}{l|}{\textbf{SWT}} & \multicolumn{1}{l|}{\textbf{Tomcat}} & \multicolumn{1}{l|}{\textbf{Overall}} \\ \hline
\multicolumn{8}{|c|}{\textbf{MRR}}                                                                                                                                                                                                                                                                                                                              \\ \hline
\multicolumn{1}{|r|}{CodeBERT}                                                      & \multicolumn{1}{r|}{0.24}             & \multicolumn{1}{r|}{0.24}          & \multicolumn{1}{r|}{0.30}             & \multicolumn{1}{r|}{0.25}         & \multicolumn{1}{r|}{0.28}         & \multicolumn{1}{r|}{0.22}            & 0.26                                  \\ \hline
\multicolumn{1}{|r|}{Long CodeBERT}                                                 & \multicolumn{1}{r|}{0.40}             & \multicolumn{1}{r|}{0.32}          & \multicolumn{1}{r|}{0.36}             & \multicolumn{1}{r|}{0.35}         & \multicolumn{1}{r|}{0.29}         & \multicolumn{1}{r|}{0.29}            & 0.34                                  \\ \hline
\multicolumn{1}{|r|}{Long RoBERTa}                                                  & \multicolumn{1}{r|}{0.31}             & \multicolumn{1}{r|}{0.31}          & \multicolumn{1}{r|}{0.33}             & \multicolumn{1}{r|}{0.35}         & \multicolumn{1}{r|}{0.33}         & \multicolumn{1}{r|}{0.32}            & 0.32                                  \\ \hline
\multicolumn{1}{|r|}{Reformer}                                                      & \multicolumn{1}{r|}{0.35}             & \multicolumn{1}{r|}{0.28}          & \multicolumn{1}{r|}{0.33}             & \multicolumn{1}{r|}{0.34}         & \multicolumn{1}{r|}{0.32}         & \multicolumn{1}{r|}{0.29}            & 0.32                                  \\ \hline
\multicolumn{8}{|c|}{\textbf{MAP}}                                                                                                                                                                                                                                                                                                                              \\ \hline
\multicolumn{1}{|r|}{CodeBERT}                                                      & \multicolumn{1}{r|}{0.16}             & \multicolumn{1}{r|}{0.16}          & \multicolumn{1}{r|}{0.21}             & \multicolumn{1}{r|}{0.16}         & \multicolumn{1}{r|}{0.19}         & \multicolumn{1}{r|}{0.13}            & 0.16                                  \\ \hline
\multicolumn{1}{|r|}{Long CodeBERT}                                                 & \multicolumn{1}{r|}{0.31}             & \multicolumn{1}{r|}{0.23}          & \multicolumn{1}{r|}{0.28}             & \multicolumn{1}{r|}{0.26}         & \multicolumn{1}{r|}{0.21}         & \multicolumn{1}{r|}{0.20}            & 0.25                                  \\ \hline
\multicolumn{1}{|r|}{Long RoBERTa}                                                  & \multicolumn{1}{r|}{0.22}             & \multicolumn{1}{r|}{0.23}          & \multicolumn{1}{r|}{0.24}             & \multicolumn{1}{r|}{0.25}         & \multicolumn{1}{r|}{0.24}         & \multicolumn{1}{r|}{0.23}            & 0.22                                  \\ \hline
\multicolumn{1}{|r|}{Reformer}                                                      & \multicolumn{1}{r|}{0.26}             & \multicolumn{1}{r|}{0.19}          & \multicolumn{1}{r|}{0.24}             & \multicolumn{1}{r|}{0.25}         & \multicolumn{1}{r|}{0.24}         & \multicolumn{1}{r|}{0.21}            & 0.22                                  \\ \hline
\end{tabular}
\label{table: average_model}
\end{table*}

\newcolumntype{s}{>{\hsize=.21\hsize}X}
\newcolumntype{d}{>{\hsize=.14\hsize}X}
\newcolumntype{v}{>{\hsize=.11\hsize}X}
\newcolumntype{k}{>{\hsize=.07\hsize}X}

\begin{table*}[!htbp]
\caption{Average performance of the embeddings in six projects sorted by sequence length.}
\label{table: average performance}
\centering
\begin{tabular}{|ccccccccc|}
\hline
\multicolumn{1}{|c|}{\textbf{\begin{tabular}[c]{@{}c@{}}Model\\ Name\end{tabular}}} & \multicolumn{1}{c|}{\textbf{\begin{tabular}[c]{@{}c@{}}Sequence\\ Length\end{tabular}}} & \multicolumn{1}{c|}{\textbf{AspectJ}} & \multicolumn{1}{c|}{\textbf{Birt}} & \multicolumn{1}{c|}{\textbf{Eclipse}} & \multicolumn{1}{c|}{\textbf{JDT}} & \multicolumn{1}{c|}{\textbf{SWT}} & \multicolumn{1}{c|}{\textbf{Tomcat}} & \textbf{Overall} \\ \hline
\multicolumn{9}{|c|}{\textbf{MRR}}                                                                                                                                                                                                                                                                                                                                                                                                   \\ \hline
\multicolumn{1}{|c|}{CodeBERT}                                                      & \multicolumn{1}{c|}{512}                                                                & \multicolumn{1}{c|}{0.24}             & \multicolumn{1}{c|}{0.24}          & \multicolumn{1}{c|}{0.30}             & \multicolumn{1}{c|}{0.25}         & \multicolumn{1}{c|}{0.28}         & \multicolumn{1}{c|}{0.22}            & 0.26             \\ \hline
\multicolumn{1}{|c|}{Long RoBERTa}                                                  & \multicolumn{1}{c|}{1536}                                                               & \multicolumn{1}{c|}{0.31}             & \multicolumn{1}{c|}{0.31}          & \multicolumn{1}{c|}{0.33}             & \multicolumn{1}{c|}{0.35}         & \multicolumn{1}{c|}{0.33}         & \multicolumn{1}{c|}{0.32}            & 0.32             \\ \hline
\multicolumn{1}{|c|}{Reformer}                                                      & \multicolumn{1}{c|}{2048}                                                               & \multicolumn{1}{c|}{0.35}             & \multicolumn{1}{c|}{0.28}          & \multicolumn{1}{c|}{0.33}             & \multicolumn{1}{c|}{0.34}         & \multicolumn{1}{c|}{0.32}         & \multicolumn{1}{c|}{0.29}            & 0.32             \\ \hline
\multicolumn{1}{|c|}{Long CodeBERT}                                                 & \multicolumn{1}{c|}{4096}                                                               & \multicolumn{1}{c|}{0.40}             & \multicolumn{1}{c|}{0.32}          & \multicolumn{1}{c|}{0.36}             & \multicolumn{1}{c|}{0.35}         & \multicolumn{1}{c|}{0.29}         & \multicolumn{1}{c|}{0.29}            & 0.34             \\ \hline
\multicolumn{9}{|c|}{\textbf{MAP}}                                                                                                                                                                                                                                                                                                                                                                                                   \\ \hline
\multicolumn{1}{|c|}{CodeBERT}                                                      & \multicolumn{1}{c|}{512}                                                                & \multicolumn{1}{c|}{0.16}             & \multicolumn{1}{c|}{0.16}          & \multicolumn{1}{c|}{0.21}             & \multicolumn{1}{c|}{0.16}         & \multicolumn{1}{c|}{0.19}         & \multicolumn{1}{c|}{0.13}            & 0.16             \\ \hline
\multicolumn{1}{|c|}{Long RoBERTa}                                                  & \multicolumn{1}{c|}{1536}                                                               & \multicolumn{1}{c|}{0.22}             & \multicolumn{1}{c|}{0.23}          & \multicolumn{1}{c|}{0.24}             & \multicolumn{1}{c|}{0.25}         & \multicolumn{1}{c|}{0.24}         & \multicolumn{1}{c|}{0.23}            & 0.22             \\ \hline
\multicolumn{1}{|c|}{Reformer}                                                      & \multicolumn{1}{c|}{2048}                                                               & \multicolumn{1}{c|}{0.26}             & \multicolumn{1}{c|}{0.19}          & \multicolumn{1}{c|}{0.24}             & \multicolumn{1}{c|}{0.25}         & \multicolumn{1}{c|}{0.24}         & \multicolumn{1}{c|}{0.21}            & 0.22             \\ \hline
\multicolumn{1}{|c|}{Long CodeBERT}                                                 & \multicolumn{1}{c|}{4096}                                                               & \multicolumn{1}{c|}{0.31}             & \multicolumn{1}{c|}{0.23}          & \multicolumn{1}{c|}{0.28}             & \multicolumn{1}{c|}{0.26}         & \multicolumn{1}{c|}{0.21}         & \multicolumn{1}{c|}{0.20}            & 0.25             \\ \hline
\end{tabular}
\end{table*}

\subsection{RQ2. Do pre-training methodologies impact embedding models performance?}
\label{RQ2}
In NLP, many pre-training methodologies are used to understand the language model. Methodologies like MLM, NSP, dynamic MLM are widely used in NLP. However, we do not know how the pre-training methodologies contribute to understanding programming language. Though several pre-training methodologies are used for natural language processing, we do not have any pre-training methods for a specific SE task (such as bug localization).\par
Moreover, Liu et al.~\citep{liu2019} have found that only some NSP pre-training is appropriate for the question-answering task in the NLP domain. Furthermore, individual sentence NSP hurts the performance of the model. However, we do not have any data about the impact of pre-training in bug localization tasks.\par
Thus it is essential to know which pre-training methodology works best for creating a joint (natural language, programming language)  embedding space and how pre-training methodologies impact embedding models performance.
Table~\ref{table: average_pretrain} shows the average performance of the bug localization model where the embedding models are trained using different pre-training methodologies. We observed that overall, \electra pre-trained embedding models produced a better result in the bug-localization task. To test the observation, we compared the performance of the models using the pairwise Mann-Whitney U test. The performance comparison between \electra,  QA methodologies was statistically significant for both MRR and MAP. For MRR, it was significant with $p=0.013$ (Bonferroni corrected $\alpha$ was $0.05/3 = 0.016$) and for MAP it was significant with $p=0.009$ (Bonferroni corrected $\alpha$ was $0.05/3 = 0.016$). However, the difference between \electra and  MLM ($p=0.3$ for MRR and $p=0.38$ for MAP)  and MLM and  QA ($p=0.97$ for MRR and $p=0.98$ for MAP) is not statistically significant. Though no specific pre-training technique performed better (statistically significant) than other techniques, \electra pre-trained embedding models achieved the highest MRR and MAP in 48\% of cases. In contrast, MLM and QA-trained embedding models achieved the highest MRR and MAP in 28\% and 23\% of cases, respectively. For CodeBERT and LongCodeBERT, we have embedding models offered by a previous study trained in a separate dataset. We have used those embedding models as it is (without training). However, the performance difference between embedding models without training and embedding models trained with \electra ($p=0.02$ for MRR and MAP) or QA ($p=0.87$ for MRR and $p=0.83$ for MAP) is not statistically significant. Only the difference between embedding models without training and models trained with MLM is statistically significant ($p=0.0001$ for both MRR and MAP). One reason for the statistically insignificant difference can be the low number of data points for comparison. As \electra pre-training is a discriminative pre-training approach, embedding models trained by this technique generate a more generalized representation. The reason for more generalization is that \electra is defined over all input tokens, whereas the MLM task is defined over a small subset of tokens. Because of the generalizability, we believe the \bl models that used \electra trained representation performed better.

\boxtext{Pre-training has an impact on the embedding model's performance.  Generally, the \electra pre-trained embedding models performed better than the other two pre-training techniques (MLM, QA) in bug localization tasks.}

\subsection{RQ3. Do transformers with longer maximum input sequence lengths perform well compared to the models with shorter maximum input sequences?}
\label{RQ3}
Before using transformers for embedding, input texts have to go through some pre-processing and tokenization steps. The text is split into tokens in the tokenization step, and a unique ID is assigned to each token. The list of IDs is called ``sequence". Typical transformers support short input sequences—for example, BERT, RoBERTa, and CodeBERT support sequences up to 512 tokens. Nevertheless, source code files are usually long. The token length distribution of the source code files in our dataset is available in the online appendix~\cite{rep_package}.
On the other hand, transformer architectures that support long input sequences require high GPU resources. The number of parameters will increase quadratically~\cite{beltagy2020longformer} with the increase of maximum sequence length. For example, the RoBERTa model has 124M trainable parameters (sequence length 512), whereas, for Reformer, it is 149M (sequence length 4096). A higher number of parameters implies that the model will require higher computing resources or longer training time (with the same computing resource). Since short input sequence transformers cannot use the complete source code, it is a common assumption that they produce low-quality embedding.\par

Ding et al.~\citep{Ding2020CogLTXAB} have studied the use of BERT for long input sequences and identified the attention decay of a typical transformer model. A typical attention mechanism requires high resources and becomes less effective over a long sequence. Thus, we may not achieve higher performance even after using higher resources.\par

Therefore, we need to know whether the cost of long input sequence transformers is justified. In this research question, we will investigate whether long input sequence transformers produce better embedding or not.
This study have used transformer models with different sequence sizes. The maximum sequence size supported by those models is presented in Table~\ref{table: average performance} along with the average performance of all the bug-localization of models' performance in six projects. We can observe that there is no trend among the models' performance except that CodeBERT performed poorly than all other models. To check the observation, we conducted a pairwise Mann-Whitney test among the models. We found that the MRR and MAP of CodeBERT model is less than the MRR and MAP of Long RoBERTa ($p<0.001$), Long CodeBERT ($p=0.002$ for MRR and $p=0.004$ for MAP) and Reformer ($p<0.001$ for both MRR and MAP) model (Bonferroni corrected $\alpha$ was $0.05/6 = 0.008$). The comparison among the other models was not statistically significant. 
The significant difference in performance among models may stem from their sequence lengths, which are 2 to 5 times longer than CodeBERT's and even 8 times longer in LongCodeBERT's case. This implies a higher computational cost for LongCodeBERT without notable performance improvements. This observation may help the developers of a bug localization tool when they have to balance between resource usage and performance gain.

\boxtext{Maximum sequence length has a mixed impact on generated embeddings' quality. The performance of transformers varies from project to project.}
\section{Related Works}
\label{sec-rel_works}
This section reviews some of the previous works related to bug localization and programming language embedding.\par

\noindent\textbf{Deep learning-based bug localization.} Recently, deep Learning approaches have been widely used to solve various software engineering problems. Some of the deep learning architectures used are Convolution Neural Networks (CNN)~\citep{mou2016convolutional, huo2016learning, lam2015combining, Li2021}, Long Short Term Memory (LSTM)~\citep{Li2019}. Lam et al.~\citep{lam2015combining} has incorporated source code metadata with the text input of a CNN-based model to achieve higher performance. DeepFL~\citep{Li2019} used a recurrent neural network followed by a multi-layer perceptron to identify faults in the Defects4J dataset. Grace~\citep{Lou2021} has used Word2Vec embedding and Gated Graph Neural Network to identify the relationship between a test case failure and the method where the fault is located. However, all of these approaches used a non-context-based embedding for their models. Moreover, the goal of these studies was to offer a better-performed model which is different from ours.

\noindent\textbf{Cross-project bug localization.} Typical bug localization models need further training if it is supposed to work on a new source-code base. Cross-project bug localization differs from typical bug localization models from this perspective. Cross-project bug localization approaches aim to create a portable bug-localization model. Zimmermann et al.~\citep{zimmermann2009cross} conducted a study on 12 projects to identify the key factors that influence the performance of a cross-project bug localization model. Moreover, they have presented a decision tress that can estimate the performance of a cross-project model based on the similarity between the training project and target project. Turhan et al.~\citep{turhan2009relative}  have used a clustering-based relevancy filtering method that groups similar data in source and target projects. The clustered data is used to train a  model, and the model's performance is tested on target models. The major drawback of this method is that it filters lots of data from the source projects. Huo et al.~\citep{Huo2019} proposed a weight-sharing approach to train a cross-project bug-localization model. Zhu et al.~\citep{Zhu2020} employed a methodology focusing solely on the Abstract Syntax Tree (AST) of source code and used an LSTM encoder for bug reports. The key distinction from other bug localization research is that their study concentrates on how different training choices affect the performance of multi-modal embeddings.
\section{Threats to Validity}
\label{sec-threats}
This section discusses potential threats to the validity of our
case studies.

\noindent\textit{\textbf{Internal Validity}.} In our study, we have trained embedding models on the BLDS dataset. For creating the BLDS dataset, we have to link bug reports with appropriate fixes.  For linking bug reports with fixes (pull requests), we have followed the approach of Liwerski et al.~\citep{liwerski2005}. However, this methodology is based on a heuristic, and a bug report might be associated with wrong pull requests and source code files. Moreover, in some cases, issues such as coding style violations are also reported as bug reports. However, to mitigate the issues, we verified the link between a bug report and pull requests by checking the pull request's attachments (if they exist). In our manual check, we found that typically non-software bug-related pull requests update many files at once. Thus we have filtered out all the pull requests that updated more than ten files. In this study, bug localization was approached as a binary classification problem using a CNN model on top of embedding models. Although ranking-based models are an option, their comparison with classification-based models requires further investigation, which is beyond the scope of this study.

\noindent\textit{\textbf{External Validity}.} A possible threat to external validity is that we evaluated the performance of the embedding on only six projects. Moreover, all of these six projects are from the same community (Apache). Thus it is possible that the performance may not represent the actual performance. However, bug reports from these six projects are often used to benchmark the performance of the bug localization model. Thus, even if the result is not generalizable, we can compare the result with other studies.

\section{Conclusion}
\label{sec-conclusion}
Our main takeaway is that the design choices made in the embeddings impact the performance of a DL similarity-based bug localization model. In this study, we have tested different design choices in creating multi-modal embedding. We have also observed the in-project and cross-project performance of the bug localization model under different settings. We have found that though large transformers require high resources, they produce better embedding for bug localization. Future research can explore creating full source code embeddings with smaller transformers and investigate the performance variance of models in in-project versus cross-project settings. Enhancing the performance of cross-project bug localization models, given the high cost of project-specific training, is also suggested as a potential research direction.


\bibliographystyle{elsarticle-num-names} 
\bibliography{Bibliography}

\begin{thebibliography}{46}
\expandafter\ifx\csname natexlab\endcsname\relax\def\natexlab#1{#1}\fi
\providecommand{\url}[1]{\texttt{#1}}
\providecommand{\href}[2]{#2}
\providecommand{\path}[1]{#1}
\providecommand{\DOIprefix}{doi:}
\providecommand{\ArXivprefix}{arXiv:}
\providecommand{\URLprefix}{URL: }
\providecommand{\Pubmedprefix}{pmid:}
\providecommand{\doi}[1]{\href{http://dx.doi.org/#1}{\path{#1}}}
\providecommand{\Pubmed}[1]{\href{pmid:#1}{\path{#1}}}
\providecommand{\bibinfo}[2]{#2}
\ifx\xfnm\relax \def\xfnm[#1]{\unskip,\space#1}\fi
\bibitem[{Saha et~al.(2013)Saha, Lease, Khurshid, and Perry}]{Saha2013}
\bibinfo{author}{R.~K. Saha}, \bibinfo{author}{M.~Lease}, \bibinfo{author}{S.~Khurshid}, \bibinfo{author}{D.~E. Perry},
\newblock \bibinfo{title}{Improving bug localization using structured information retrieval},
\newblock in: \bibinfo{booktitle}{2013 28th {IEEE}/{ACM} International Conference on Automated Software Engineering ({ASE})}, \bibinfo{publisher}{{IEEE}}, \bibinfo{year}{2013}.
\bibitem[{Wang and Lo(2014)}]{Wang2014}
\bibinfo{author}{S.~Wang}, \bibinfo{author}{D.~Lo},
\newblock \bibinfo{title}{Version history, similar report, and structure: putting them together for improved bug localization},
\newblock in: \bibinfo{booktitle}{Proceedings of the 22nd International Conference on Program Comprehension - {ICPC} 2014}, \bibinfo{publisher}{{ACM} Press}, \bibinfo{year}{2014}.
\bibitem[{Wen et~al.(2016)Wen, Wu, and Cheung}]{Wen2016}
\bibinfo{author}{M.~Wen}, \bibinfo{author}{R.~Wu}, \bibinfo{author}{S.-C. Cheung},
\newblock \bibinfo{title}{Locus: locating bugs from software changes},
\newblock in: \bibinfo{booktitle}{Proceedings of the 31st {IEEE}/{ACM} International Conference on Automated Software Engineering}, \bibinfo{publisher}{{ACM}}, \bibinfo{year}{2016}.
\bibitem[{Wong et~al.(2014)Wong, Xiong, Zhang, Hao, Zhang, and Mei}]{Wong2014}
\bibinfo{author}{C.-P. Wong}, \bibinfo{author}{Y.~Xiong}, \bibinfo{author}{H.~Zhang}, \bibinfo{author}{D.~Hao}, \bibinfo{author}{L.~Zhang}, \bibinfo{author}{H.~Mei},
\newblock \bibinfo{title}{Boosting bug-report-oriented fault localization with segmentation and stack-trace analysis},
\newblock in: \bibinfo{booktitle}{2014 {IEEE} International Conference on Software Maintenance and Evolution}, \bibinfo{publisher}{{IEEE}}, \bibinfo{year}{2014}.
\bibitem[{Youm et~al.(2015)Youm, Ahn, Kim, and Lee}]{Youm2015}
\bibinfo{author}{K.~C. Youm}, \bibinfo{author}{J.~Ahn}, \bibinfo{author}{J.~Kim}, \bibinfo{author}{E.~Lee},
\newblock \bibinfo{title}{Bug localization based on code change histories and bug reports},
\newblock in: \bibinfo{booktitle}{2015 Asia-Pacific Software Engineering Conference ({APSEC})}, \bibinfo{publisher}{{IEEE}}, \bibinfo{year}{2015}.
\bibitem[{Zhou et~al.(2012)Zhou, Zhang, and Lo}]{Zhou2012}
\bibinfo{author}{J.~Zhou}, \bibinfo{author}{H.~Zhang}, \bibinfo{author}{D.~Lo},
\newblock \bibinfo{title}{Where should the bugs be fixed? more accurate information retrieval-based bug localization based on bug reports},
\newblock in: \bibinfo{booktitle}{2012 34th International Conference on Software Engineering ({ICSE})}, \bibinfo{publisher}{{IEEE}}, \bibinfo{year}{2012}.
\bibitem[{Akbar and Kak(2020)}]{Akbar2020}
\bibinfo{author}{S.~A. Akbar}, \bibinfo{author}{A.~C. Kak},
\newblock \bibinfo{title}{A large-scale comparative evaluation of {IR}-based tools for bug localization},
\newblock in: \bibinfo{booktitle}{Proceedings of the 17th International Conference on Mining Software Repositories}, \bibinfo{publisher}{{ACM}}, \bibinfo{year}{2020}.
\bibitem[{Moin and Khansari(2010)}]{Moin2010}
\bibinfo{author}{A.~H. Moin}, \bibinfo{author}{M.~Khansari},
\newblock \bibinfo{title}{Bug localization using revision log analysis and open bug repository text categorization},
\newblock in: \bibinfo{booktitle}{{IFIP} Advances in Information and Communication Technology}, \bibinfo{publisher}{Springer Berlin Heidelberg}, \bibinfo{year}{2010}, pp. \bibinfo{pages}{188--199}.
\bibitem[{Liu et~al.(2005)Liu, Yan, Fei, Han, and Midkiff}]{Liu2005}
\bibinfo{author}{C.~Liu}, \bibinfo{author}{X.~Yan}, \bibinfo{author}{L.~Fei}, \bibinfo{author}{J.~Han}, \bibinfo{author}{S.~P. Midkiff},
\newblock \bibinfo{title}{{SOBER}},
\newblock \bibinfo{journal}{{ACM} {SIGSOFT} Software Engineering Notes} \bibinfo{volume}{30} (\bibinfo{year}{2005}) \bibinfo{pages}{286--295}.
\bibitem[{Le et~al.(2015)Le, Oentaryo, and Lo}]{Le2015}
\bibinfo{author}{T.-D.~B. Le}, \bibinfo{author}{R.~J. Oentaryo}, \bibinfo{author}{D.~Lo},
\newblock \bibinfo{title}{Information retrieval and spectrum based bug localization: better together},
\newblock in: \bibinfo{booktitle}{Proceedings of the 2015 10th Joint Meeting on Foundations of Software Engineering}, \bibinfo{publisher}{{ACM}}, \bibinfo{year}{2015}.
\bibitem[{Dallmeier et~al.(2005)Dallmeier, Lindig, and Zeller}]{Dallmeier2005}
\bibinfo{author}{V.~Dallmeier}, \bibinfo{author}{C.~Lindig}, \bibinfo{author}{A.~Zeller},
\newblock \bibinfo{title}{Lightweight bug localization with {AMPLE}},
\newblock in: \bibinfo{booktitle}{Proceedings of the Sixth sixth international symposium on Automated analysis-driven debugging - {AADEBUG}{\textquotesingle}05}, \bibinfo{publisher}{{ACM} Press}, \bibinfo{year}{2005}.
\bibitem[{Huo et~al.(2019)Huo, Thung, Li, Lo, and Shi}]{Huo2019}
\bibinfo{author}{X.~Huo}, \bibinfo{author}{F.~Thung}, \bibinfo{author}{M.~Li}, \bibinfo{author}{D.~Lo}, \bibinfo{author}{S.-T. Shi},
\newblock \bibinfo{title}{Deep transfer bug localization},
\newblock \bibinfo{journal}{{IEEE} Transactions on Software Engineering}  (\bibinfo{year}{2019}) \bibinfo{pages}{1--1}.
\bibitem[{Huo et~al.(2016)Huo, Li, Zhou et~al.}]{huo2016learning}
\bibinfo{author}{X.~Huo}, \bibinfo{author}{M.~Li}, \bibinfo{author}{Z.-H. Zhou}, et~al.,
\newblock \bibinfo{title}{Learning unified features from natural and programming languages for locating buggy source code.},
\newblock in: \bibinfo{booktitle}{IJCAI}, volume~\bibinfo{volume}{16}, \bibinfo{year}{2016}, pp. \bibinfo{pages}{1606--1612}.
\bibitem[{Wang et~al.(2020)Wang, Xu, Yan, Liu, and Liu}]{Wang2020}
\bibinfo{author}{B.~Wang}, \bibinfo{author}{L.~Xu}, \bibinfo{author}{M.~Yan}, \bibinfo{author}{C.~Liu}, \bibinfo{author}{L.~Liu},
\newblock \bibinfo{title}{Multi-dimension convolutional neural network for bug localization},
\newblock \bibinfo{journal}{{IEEE} Transactions on Services Computing}  (\bibinfo{year}{2020}) \bibinfo{pages}{1--1}.
\bibitem[{Jiang et~al.(2020)Jiang, Liu, and Xu}]{Jiang2020}
\bibinfo{author}{B.~Jiang}, \bibinfo{author}{P.~Liu}, \bibinfo{author}{J.~Xu},
\newblock \bibinfo{title}{A deep learning approach to locate buggy files},
\newblock in: \bibinfo{booktitle}{2020 {IEEE} 11th International Conference on Dependable Systems, Services and Technologies ({DESSERT})}, \bibinfo{publisher}{{IEEE}}, \bibinfo{year}{2020}.
\bibitem[{Zhu et~al.(2020)Zhu, Li, Tong, and Wang}]{Zhu2020}
\bibinfo{author}{Z.~Zhu}, \bibinfo{author}{Y.~Li}, \bibinfo{author}{H.~Tong}, \bibinfo{author}{Y.~Wang},
\newblock \bibinfo{title}{{CooBa}: Cross-project bug localization via adversarial transfer learning},
\newblock in: \bibinfo{booktitle}{Proceedings of the Twenty-Ninth International Joint Conference on Artificial Intelligence}, \bibinfo{publisher}{International Joint Conferences on Artificial Intelligence Organization}, \bibinfo{year}{2020}.
\bibitem[{Xiao et~al.(2018)Xiao, Keung, Mi, and Bennin}]{Xiao2018}
\bibinfo{author}{Y.~Xiao}, \bibinfo{author}{J.~Keung}, \bibinfo{author}{Q.~Mi}, \bibinfo{author}{K.~E. Bennin},
\newblock \bibinfo{title}{Bug localization with semantic and structural features using convolutional neural network and cascade forest},
\newblock in: \bibinfo{booktitle}{Proceedings of the 22nd International Conference on Evaluation and Assessment in Software Engineering 2018}, \bibinfo{publisher}{{ACM}}, \bibinfo{year}{2018}.
\bibitem[{Devlin et~al.(2018)Devlin, Chang, Lee, and Toutanova}]{devlin2018bert}
\bibinfo{author}{J.~Devlin}, \bibinfo{author}{M.-W. Chang}, \bibinfo{author}{K.~Lee}, \bibinfo{author}{K.~Toutanova},
\newblock \bibinfo{title}{Bert: Pre-training of deep bidirectional transformers for language understanding},
\newblock \bibinfo{journal}{arXiv preprint arXiv:1810.04805}  (\bibinfo{year}{2018}).
\bibitem[{Liu et~al.(2019)Liu, Ott, Goyal, Du, Joshi, Chen, Levy, Lewis, Zettlemoyer, and Stoyanov}]{liu2019}
\bibinfo{author}{Y.~Liu}, \bibinfo{author}{M.~Ott}, \bibinfo{author}{N.~Goyal}, \bibinfo{author}{J.~Du}, \bibinfo{author}{M.~Joshi}, \bibinfo{author}{D.~Chen}, \bibinfo{author}{O.~Levy}, \bibinfo{author}{M.~Lewis}, \bibinfo{author}{L.~Zettlemoyer}, \bibinfo{author}{V.~Stoyanov}, \bibinfo{title}{Roberta: A robustly optimized bert pretraining approach}, \bibinfo{year}{2019}. \href{http://arxiv.org/abs/1907.11692}{{\tt arXiv:1907.11692}}.
\bibitem[{Onta{\~{n}}{\'{o}}n et~al.(2021)Onta{\~{n}}{\'{o}}n, Ainslie, Cvicek, and Fisher}]{Santiago2021}
\bibinfo{author}{S.~Onta{\~{n}}{\'{o}}n}, \bibinfo{author}{J.~Ainslie}, \bibinfo{author}{V.~Cvicek}, \bibinfo{author}{Z.~Fisher},
\newblock \bibinfo{title}{Making transformers solve compositional tasks},
\newblock \bibinfo{journal}{CoRR} \bibinfo{volume}{abs/2108.04378} (\bibinfo{year}{2021}).
\bibitem[{Zhu et~al.(2021)Zhu, Wang, Ni, and Xie}]{Zhu2021}
\bibinfo{author}{W.~Zhu}, \bibinfo{author}{X.~Wang}, \bibinfo{author}{Y.~Ni}, \bibinfo{author}{G.~Xie},
\newblock \bibinfo{title}{{AutoTrans}: Automating transformer design via reinforced architecture search},
\newblock in: \bibinfo{booktitle}{Natural Language Processing and Chinese Computing}, \bibinfo{publisher}{Springer International Publishing}, \bibinfo{year}{2021}, pp. \bibinfo{pages}{169--182}.
\bibitem[{Izsak et~al.(2021)Izsak, Berchansky, and Levy}]{Izsak2021}
\bibinfo{author}{P.~Izsak}, \bibinfo{author}{M.~Berchansky}, \bibinfo{author}{O.~Levy},
\newblock \bibinfo{title}{How to train {BERT} with an academic budget},
\newblock \bibinfo{journal}{CoRR} \bibinfo{volume}{abs/2104.07705} (\bibinfo{year}{2021}). \href{http://arxiv.org/abs/2104.07705}{{\tt arXiv:2104.07705}}.
\bibitem[{Quijano et~al.(2021)Quijano, Nguyen, and Ordonez}]{Quijano2021}
\bibinfo{author}{A.~J. Quijano}, \bibinfo{author}{S.~Nguyen}, \bibinfo{author}{J.~Ordonez},
\newblock \bibinfo{title}{Grid search hyperparameter benchmarking of bert, albert, and longformer on duorc},
\newblock \bibinfo{journal}{CoRR} \bibinfo{volume}{abs/2101.06326} (\bibinfo{year}{2021}). \href{http://arxiv.org/abs/2101.06326}{{\tt arXiv:2101.06326}}.
\bibitem[{Liang et~al.(2022)Liang, Hang, and Li}]{Liang2022}
\bibinfo{author}{H.~Liang}, \bibinfo{author}{D.~Hang}, \bibinfo{author}{X.~Li},
\newblock \bibinfo{title}{Modeling function-level interactions for file-level bug localization},
\newblock \bibinfo{journal}{Empirical Software Engineering} \bibinfo{volume}{27} (\bibinfo{year}{2022}).
\bibitem[{Peters et~al.(2019)Peters, Ruder, and Smith}]{peters-etal-2019-tune}
\bibinfo{author}{M.~E. Peters}, \bibinfo{author}{S.~Ruder}, \bibinfo{author}{N.~A. Smith},
\newblock \bibinfo{title}{To tune or not to tune? adapting pretrained representations to diverse tasks},
\newblock in: \bibinfo{booktitle}{Proceedings of the 4th Workshop on Representation Learning for NLP (RepL4NLP-2019)}, \bibinfo{publisher}{Association for Computational Linguistics}, \bibinfo{address}{Florence, Italy}, \bibinfo{year}{2019}, pp. \bibinfo{pages}{7--14}.
\bibitem[{Beltagy et~al.(2020)Beltagy, Peters, and Cohan}]{beltagy2020longformer}
\bibinfo{author}{I.~Beltagy}, \bibinfo{author}{M.~E. Peters}, \bibinfo{author}{A.~Cohan},
\newblock \bibinfo{title}{Longformer: The long-document transformer},
\newblock \bibinfo{journal}{arXiv preprint arXiv:2004.05150}  (\bibinfo{year}{2020}).
\bibitem[{Ye et~al.(2014)Ye, Bunescu, and Liu}]{Ye2014}
\bibinfo{author}{X.~Ye}, \bibinfo{author}{R.~Bunescu}, \bibinfo{author}{C.~Liu},
\newblock \bibinfo{title}{Learning to rank relevant files for bug reports using domain knowledge},
\newblock in: \bibinfo{booktitle}{Proceedings of the 22nd {ACM} {SIGSOFT} International Symposium on Foundations of Software Engineering - {FSE} 2014}, \bibinfo{publisher}{{ACM} Press}, \bibinfo{year}{2014}.
\bibitem[{{\'{S}}liwerski et~al.(2005){\'{S}}liwerski, Zimmermann, and Zeller}]{liwerski2005}
\bibinfo{author}{J.~{\'{S}}liwerski}, \bibinfo{author}{T.~Zimmermann}, \bibinfo{author}{A.~Zeller},
\newblock \bibinfo{title}{When do changes induce fixes?},
\newblock \bibinfo{journal}{{ACM} {SIGSOFT} Software Engineering Notes} \bibinfo{volume}{30} (\bibinfo{year}{2005}) \bibinfo{pages}{1--5}.
\bibitem[{Ye(2014)}]{Ye2014_dataset}
\bibinfo{author}{X.~Ye},
\newblock \bibinfo{title}{{The dataset of six open source Java projects}}  (\bibinfo{year}{2014}). \URLprefix \url{https://figshare.com/articles/dataset/The_dataset_of_six_open_source_Java_projects/951967}. \DOIprefix\doi{10.6084/m9.figshare.951967.v10}.
\bibitem[{Kim and Lee(2021)}]{Kim2021}
\bibinfo{author}{M.~Kim}, \bibinfo{author}{E.~Lee},
\newblock \bibinfo{title}{Are datasets for information retrieval-based bug localization techniques trustworthy?},
\newblock \bibinfo{journal}{Empirical Software Engineering} \bibinfo{volume}{26} (\bibinfo{year}{2021}).
\bibitem[{Lee et~al.(2018)Lee, Kim, Bissyand{\'{e}}, Jung, and Traon}]{Lee2018}
\bibinfo{author}{J.~Lee}, \bibinfo{author}{D.~Kim}, \bibinfo{author}{T.~F. Bissyand{\'{e}}}, \bibinfo{author}{W.~Jung}, \bibinfo{author}{Y.~L. Traon},
\newblock \bibinfo{title}{Bench4bl: reproducibility study on the performance of {IR}-based bug localization},
\newblock in: \bibinfo{booktitle}{Proceedings of the 27th {ACM} {SIGSOFT} International Symposium on Software Testing and Analysis}, \bibinfo{publisher}{{ACM}}, \bibinfo{year}{2018}.
\bibitem[{Moreno et~al.(2015)Moreno, Bavota, Haiduc, Penta, Oliveto, Russo, and Marcus}]{Moreno2015}
\bibinfo{author}{L.~Moreno}, \bibinfo{author}{G.~Bavota}, \bibinfo{author}{S.~Haiduc}, \bibinfo{author}{M.~D. Penta}, \bibinfo{author}{R.~Oliveto}, \bibinfo{author}{B.~Russo}, \bibinfo{author}{A.~Marcus},
\newblock \bibinfo{title}{Query-based configuration of text retrieval solutions for software engineering tasks},
\newblock in: \bibinfo{booktitle}{Proceedings of the 2015 10th Joint Meeting on Foundations of Software Engineering}, \bibinfo{publisher}{{ACM}}, \bibinfo{year}{2015}.
\bibitem[{Feng et~al.(2020)Feng, Guo, Tang, Duan, Feng, Gong, Shou, Qin, Liu, Jiang, and Zhou}]{Feng2020}
\bibinfo{author}{Z.~Feng}, \bibinfo{author}{D.~Guo}, \bibinfo{author}{D.~Tang}, \bibinfo{author}{N.~Duan}, \bibinfo{author}{X.~Feng}, \bibinfo{author}{M.~Gong}, \bibinfo{author}{L.~Shou}, \bibinfo{author}{B.~Qin}, \bibinfo{author}{T.~Liu}, \bibinfo{author}{D.~Jiang}, \bibinfo{author}{M.~Zhou},
\newblock \bibinfo{title}{{CodeBERT}: A pre-trained model for programming and natural languages},
\newblock in: \bibinfo{booktitle}{Findings of the Association for Computational Linguistics: {EMNLP} 2020}, \bibinfo{publisher}{Association for Computational Linguistics}, \bibinfo{year}{2020}.
\bibitem[{Clark et~al.(2020)Clark, Luong, Le, and Manning}]{clark2020electra}
\bibinfo{author}{K.~Clark}, \bibinfo{author}{M.-T. Luong}, \bibinfo{author}{Q.~V. Le}, \bibinfo{author}{C.~D. Manning},
\newblock \bibinfo{title}{Electra: Pre-training text encoders as discriminators rather than generators},
\newblock \bibinfo{journal}{arXiv preprint arXiv:2003.10555}  (\bibinfo{year}{2020}).
\bibitem[{Glass et~al.(2020)Glass, Gliozzo, Chakravarti, Ferritto, Pan, Bhargav, Garg, and Sil}]{Glass2020}
\bibinfo{author}{M.~Glass}, \bibinfo{author}{A.~Gliozzo}, \bibinfo{author}{R.~Chakravarti}, \bibinfo{author}{A.~Ferritto}, \bibinfo{author}{L.~Pan}, \bibinfo{author}{G.~P.~S. Bhargav}, \bibinfo{author}{D.~Garg}, \bibinfo{author}{A.~Sil},
\newblock \bibinfo{title}{Span selection pre-training for question answering},
\newblock in: \bibinfo{booktitle}{Proceedings of the 58th Annual Meeting of the Association for Computational Linguistics}, \bibinfo{publisher}{Association for Computational Linguistics}, \bibinfo{address}{Online}, \bibinfo{year}{2020}, pp. \bibinfo{pages}{2773--2782}.
\bibitem[{Huo et~al.(2021)Huo, Thung, Li, Lo, and Shi}]{Huo2021}
\bibinfo{author}{X.~Huo}, \bibinfo{author}{F.~Thung}, \bibinfo{author}{M.~Li}, \bibinfo{author}{D.~Lo}, \bibinfo{author}{S.-T. Shi},
\newblock \bibinfo{title}{Deep transfer bug localization},
\newblock \bibinfo{journal}{{IEEE} Transactions on Software Engineering} \bibinfo{volume}{47} (\bibinfo{year}{2021}) \bibinfo{pages}{1368--1380}.
\bibitem[{Anonymous(2024)}]{rep_package}
\bibinfo{author}{Anonymous},
\newblock \bibinfo{title}{Aligning programming language and natural language: Exploring design choices in multi-modal transformer-based embedding for bug localization}  (\bibinfo{year}{2024}). \URLprefix \url{https://zenodo.org/doi/10.5281/zenodo.6760333}. \DOIprefix\doi{10.5281/ZENODO.6760333}.
\bibitem[{Howard and Ruder(2018)}]{howard2018}
\bibinfo{author}{J.~Howard}, \bibinfo{author}{S.~Ruder},
\newblock \bibinfo{title}{Universal language model fine-tuning for text classification},
\newblock in: \bibinfo{booktitle}{Proceedings of the 56th Annual Meeting of the Association for Computational Linguistics (Volume 1: Long Papers)}, \bibinfo{publisher}{Association for Computational Linguistics}, \bibinfo{address}{Melbourne, Australia}, \bibinfo{year}{2018}, pp. \bibinfo{pages}{328--339}.
\bibitem[{Ding et~al.(2020)Ding, Zhou, Yang, and Tang}]{Ding2020CogLTXAB}
\bibinfo{author}{M.~Ding}, \bibinfo{author}{C.~Zhou}, \bibinfo{author}{H.~Yang}, \bibinfo{author}{J.~Tang},
\newblock \bibinfo{title}{Cogltx: Applying bert to long texts},
\newblock in: \bibinfo{booktitle}{NeurIPS}, \bibinfo{year}{2020}.
\bibitem[{Mou et~al.(2016)Mou, Li, Zhang, Wang, and Jin}]{mou2016convolutional}
\bibinfo{author}{L.~Mou}, \bibinfo{author}{G.~Li}, \bibinfo{author}{L.~Zhang}, \bibinfo{author}{T.~Wang}, \bibinfo{author}{Z.~Jin},
\newblock \bibinfo{title}{Convolutional neural networks over tree structures for programming language processing},
\newblock in: \bibinfo{booktitle}{Proceedings of the AAAI Conference on Artificial Intelligence}, volume~\bibinfo{volume}{30}, \bibinfo{year}{2016}.
\bibitem[{Lam et~al.(2015)Lam, Nguyen, Nguyen, and Nguyen}]{lam2015combining}
\bibinfo{author}{A.~N. Lam}, \bibinfo{author}{A.~T. Nguyen}, \bibinfo{author}{H.~A. Nguyen}, \bibinfo{author}{T.~N. Nguyen},
\newblock \bibinfo{title}{Combining deep learning with information retrieval to localize buggy files for bug reports (n)},
\newblock in: \bibinfo{booktitle}{2015 30th IEEE/ACM International Conference on Automated Software Engineering (ASE)}, \bibinfo{organization}{IEEE}, \bibinfo{year}{2015}, pp. \bibinfo{pages}{476--481}.
\bibitem[{Li et~al.(2021)Li, Wang, and Nguyen}]{Li2021}
\bibinfo{author}{Y.~Li}, \bibinfo{author}{S.~Wang}, \bibinfo{author}{T.~Nguyen},
\newblock \bibinfo{title}{Fault localization with code coverage representation learning},
\newblock in: \bibinfo{booktitle}{2021 {IEEE}/{ACM} 43rd International Conference on Software Engineering ({ICSE})}, \bibinfo{publisher}{{IEEE}}, \bibinfo{year}{2021}.
\bibitem[{Li et~al.(2019)Li, Li, Zhang, and Zhang}]{Li2019}
\bibinfo{author}{X.~Li}, \bibinfo{author}{W.~Li}, \bibinfo{author}{Y.~Zhang}, \bibinfo{author}{L.~Zhang},
\newblock \bibinfo{title}{{DeepFL}: integrating multiple fault diagnosis dimensions for deep fault localization},
\newblock in: \bibinfo{booktitle}{Proceedings of the 28th {ACM} {SIGSOFT} International Symposium on Software Testing and Analysis}, \bibinfo{publisher}{{ACM}}, \bibinfo{year}{2019}.
\bibitem[{Lou et~al.(2021)Lou, Zhu, Dong, Li, Sun, Hao, Zhang, and Zhang}]{Lou2021}
\bibinfo{author}{Y.~Lou}, \bibinfo{author}{Q.~Zhu}, \bibinfo{author}{J.~Dong}, \bibinfo{author}{X.~Li}, \bibinfo{author}{Z.~Sun}, \bibinfo{author}{D.~Hao}, \bibinfo{author}{L.~Zhang}, \bibinfo{author}{L.~Zhang},
\newblock \bibinfo{title}{Boosting coverage-based fault localization via graph-based representation learning},
\newblock in: \bibinfo{booktitle}{Proceedings of the 29th {ACM} Joint Meeting on European Software Engineering Conference and Symposium on the Foundations of Software Engineering}, \bibinfo{publisher}{{ACM}}, \bibinfo{year}{2021}.
\bibitem[{Zimmermann et~al.(2009)Zimmermann, Nagappan, Gall, Giger, and Murphy}]{zimmermann2009cross}
\bibinfo{author}{T.~Zimmermann}, \bibinfo{author}{N.~Nagappan}, \bibinfo{author}{H.~Gall}, \bibinfo{author}{E.~Giger}, \bibinfo{author}{B.~Murphy},
\newblock \bibinfo{title}{Cross-project defect prediction: a large scale experiment on data vs. domain vs. process},
\newblock in: \bibinfo{booktitle}{Proceedings of the 7th joint meeting of the European software engineering conference and the ACM SIGSOFT symposium on The foundations of software engineering}, \bibinfo{year}{2009}, pp. \bibinfo{pages}{91--100}.
\bibitem[{Turhan et~al.(2009)Turhan, Menzies, Bener, and Di~Stefano}]{turhan2009relative}
\bibinfo{author}{B.~Turhan}, \bibinfo{author}{T.~Menzies}, \bibinfo{author}{A.~B. Bener}, \bibinfo{author}{J.~Di~Stefano},
\newblock \bibinfo{title}{On the relative value of cross-company and within-company data for defect prediction},
\newblock \bibinfo{journal}{Empirical Software Engineering} \bibinfo{volume}{14} (\bibinfo{year}{2009}) \bibinfo{pages}{540--578}.

\end{thebibliography}

\end{document}